# $Z_2$ GAUGE THEORY OF ELECTRON FRACTIONALIZATION IN THE $t, t' - J$ MODEL WITH UNIAXIAL ANISOTROPY

D. Schmeltzer
Dept. of Physics City College
Of the City University of New York
New York, NY  10031
and
A. R. Bishop
Theoretical Division
Los Alamos National Lab.
Los Alamos, NM  87545

Abstract

We consider a strongly correlated "$t, t' - J$" spin model with a positive uniaxial anisotropy.  We show that in $2 + 1$ dimensions, this model is equivalent to spinons and holons  excitations coupled to  a $Z_2$ Ising gauge field. The  $Z_2$ gauge field represent the vortex monopoles excitations of the X-Y model.  As a function of the holes concentrations the X-Y model has two phases: a spin wave phase  and  a free vortex monopole phase. In the spin wave phase the spinons and the holons are weakly  interacting giving rise to a spin –charge separated phase  which at zero temperature is is superconducting. When the holes concentrations increases   the $Z_2$  excitations (the monopolesl currents ) become free giving rise to  holon-spinon binding . The destruction of superconductivity at zero temperature coincides with the appearance of the  free  monopoles currents.

1.INTRODUCTION

One of the fundamental questions in the high $T_c$ cuprate superconductors is the possibility of spin-charge separation in $2 + 1$ dimensions. Following Anderson (1), the high $T_c$ superconductors have been described within the $t - J$ model. The $t - J$ model has been shown to be equivalent to a liquid of charged holes coupled to spin excitations. Anderson has argued that the new excitations of the systems are the charged holons ( the fermionic holes ) and neutral spinons (the disorder spin liquid) which interact trough a $U(1)$ gauge field (2,3).The recent angle –resolved photoemission spectroscopy (13) and optical conductivity (14) support Anderson 's picture .This picture points towards a spin charge separation model ,a model which is well supported experimentally. Theoretically the picture of charge and spin –separation is equivalent to deconfinement .The regular electron represents the confinement phase of holons and spinons mediated by the U(1) gauge field. According to Fradkin and Shenker (4) the U(1) gauge theory in $2 + 1$ dimensions is always in the confined phase (5). On the other hand, if the U(1) gauge field is replaced by a $Z_2$ Ising gauge field the model has a deconfined phase. This idea has been exploited by Senthil and Fisher (6) who have argued that the $t - J$ model is equivalent to holons and spinons interacting with a $Z_2$ gauge field and therefore they concluded that spin charge separation occurs. The argument presented by Senthil and Fisher (6) has been criticized by Hastings (7) who has shown that the $Z_2$ symmetry is a result of an approximation .When the exact effective action is constructed one recovers the full U(1) sgauge symmetry. Therefore we have only a confined phase and no spin – charge separation!

In order to allow for a deconfined phase we propose to modify the model by introducing a positive uniaxial magnetic anisotropy. Such a term originates from the lattice crystal field effects. (The presence of the uniaxial anisotropy term replaces the Heisenberg model by an X-Y model (8) and the presence of magnetic frustrations (9-11) gives rise to a non-magnetic ground state.)

The purpose of this paper is to present a theoretical formulation which is consistent with the experimental observation where for the underdoped case a spin charge separated state appears. The magnetic order is destroyed and the ground state at zero temperature is superconducting. At finite temperatures one obtains a spin charge separated metal coined non-fermi liquid. At large doping superconductivity is suppressed and a regular metal emerges.

Formally we achieve this by adding an uniaxial term to the Hamiltonian which breaks the SU(2) symmetry. The "uniaxial" term is given by $\frac{1}{2}D\sum_{\vec{r}} n_s^2(\vec{r})$, $D > 0$, and $n_s(\vec{r}) = n_\uparrow(\vec{r}) - n_\downarrow(\vec{r})$. As a result the Schwinger boson representation for the spin is replaced by an U(1) spinon representation.

When the system is doped the the (Ising) magnetic order $n_S(\vec{r})$ vanishes. The holes doping is described by the spinless fermions density $\psi^+(\vec{r})\psi(\vec{r})$. In the presence of holes the magnetic magnetic order, $n_s(\vec{r}) = C_\uparrow^+(\vec{r})C_\uparrow(\vec{r}) - C_\downarrow^+(\vec{r})C_\downarrow(\vec{r})$, gives rise to a constraint at each site, $n_s(\vec{r}) = \pm\left(1 - \psi^+(\vec{r})\psi(\vec{r})\right)$. (The meaning of this constrained is that in the absence of holes the magnetization "$n_s(\vec{r})$" in the the "z" direction takes values $\pm 1$, when the holes density is one the magnetization vanishes.) At finite doping $\delta > \delta_c$ the Ising magnetic order vanishes. For this condition the spin excitations are described by neutral spinons characterized by the bosonic phase .The major cosequence of this representation is that the fermionic holons ( the holes with no spin) interact with the bosonic spinons . This interaction is given by a real order hopping term which replaces the U(1) gauge field, see eqs.6a-6b in the paper. It is this real hopping term which is the source of the $Z_2$ symmetry. The spinon bosonic phase excitations are controlled by the X-Y hamiltonian which replaces the Heisenberg model ( the Berry phase of the modified Heisenberg model in 2+1 dimensions vanishes). The X-Y model has an effective coupling constant $\gamma_{eff}$ .This coupling constant is the effective X-Y exchange coupling constant which decreases with the increase of the holes concentrations $\delta$ . In this regime the $Z_2$ excitations represents the vortex monopoles currents in of the X-Y model in 2+1 dimensions. The $X-Y$ model in $2+1$ dimensions has a phase transition at a coupling constant $\gamma_{eff} = \gamma^*$, where $\gamma^* \equiv \gamma(\delta^*)$ and $\delta^*$ is the quantum critical holes concentrations ($\delta_c < \delta^*$ ). For $\gamma_{eff} > \gamma^*$ the $X-Y$ model is in the spin wave phase with no monopole currents .In this region the spinons and the holons are weakly interacting giving rise to spin- charge separation. At zero temperature the ground state is superconducting . This phase occurs at low holes concentrations , $\delta_c < \delta < \delta^*$. At large holes concentrations , $\delta^* < \delta$ the coupling constant obeys, $\gamma_{eff} < \gamma^*$. As a result the monopoles currents are free ,spin and charge confinement takes place and superconductivity is destroyed at zero temperature. We identify the critical value $\gamma^* = \gamma(\delta^*)$ wih the quantum critical point.

The plane of this paper is as followings: section 2 is devoted to the presentation of the model ,in sections 3 and 4 we solve the constraints problem and find the action for the fermionic charges (holons) and bosonic spin degrees of freedom ( spinons in the spin-waves phase or monopoles currents) ,in section 5 we construct the Euclidean action ,we obtain a bosonic Berry phase for the spinon ( which is different from the Heisenberg Berry phase), section 6 is restricted to the situation where the uniaxial term gives rise to a non-magnetic ground state, in section 7 we consider the phase transition of the X-Y model in 2+1 dimension as a function of the holes concentrations and show that the spin charge separated phase occurs in the spinon-spin wave phase (the phase with no monopole currents) and in section 8 we present our conclusions.

2. THE MODEL

We consider the following model the t –t'-J model with uniaxial anisotropy.

$$
\begin{aligned}
H = &-t \sum_{\vec{r},i=1,2} \sum_{\sigma=\uparrow,\downarrow} C^+_\sigma(\vec{r}) C_\sigma(\vec{r}+a_i) + \text{H.C.} \\
&-t' \sum_{\vec{r},\pm} \sum_{\sigma=\uparrow,\downarrow} C^+_\sigma(\vec{r}) C_\sigma(\vec{r}+a_1 \pm a_2) + \text{H.C.} \\
&+ J \sum_{\vec{r},i=1,2} \left[ \vec{S}(\vec{r}) \cdot \vec{S}(\vec{r}+a_i) - \frac{1}{4} n(\vec{r}) n(\vec{r}+a_i) \right] \\
&+ \frac{D}{2} \sum_{\vec{r}} n_s^2(\vec{r})
\end{aligned}
\tag{1}
$$

where, $n_s(\vec{r}) = C^+_\uparrow(\vec{r}) C_\uparrow(\vec{r}) - C^+_\downarrow(\vec{r}) C_\downarrow(\vec{r})$ is the magnetization , $D > 0$ represents the uniaxial anisotropy,"t" and "t'" are the nearest and next nearest neighbor hopping, and "J" is the Heisenberg exchange coupling. In eq. (1), $\vec{S}(r)$ is the spin-1/2 density, and $n(\vec{r})$ is the charge density. As $D \to 0$, $t \to 0$ and $\delta \to 0$ (no doping) the model in eq. (1)

describes an Heisenberg antiferromagnet. When the uniaxial field is large, D>>0 the Heisenberg model is replaced by an X-Y model in 2+1 dimensions .The excitations of the X-Y model (spinons-spin waves and monopoles currents) play a crucial role in our investigation.

3. THE BOSONIC "DENSITY –PHASE" REPRESENTATION OF THE SPIN EXCITATIONS

In the remainder of this paper, we will present details of our calculations. We will show that the spin degrees of freedom can be represented in terms of the bosonic "density-phase " representations.
. We will show that the fermionic holons excitations interact with the bosonic phase of the spinon excitations.Our starting point is the slave boson representation:

$$C_\sigma^+ = b_\sigma^+(\vec{r})\widetilde{\psi}(\vec{r}), C_\sigma(\vec{r}) = \widetilde{\psi}_\sigma^+ b_\sigma(\vec{r}) \qquad (2)$$

with the constraint,

$$\widetilde{\psi}^+(\vec{r})\widetilde{\psi}(\vec{r}) + b_\uparrow^+(\vec{r})b_\uparrow(\vec{r}) + b_\downarrow^+(\vec{r})b_\downarrow(\vec{r}) = 1 \quad , \qquad (3)$$

where $\widetilde{\psi}^+, \widetilde{\psi}$ are spinless fermions which create and destroy a holons . $b_\sigma^+$, $b_\sigma$, $\sigma = \uparrow,\downarrow$ represent the Schwinger bosononic representation for spin spin ,creation and annihilation. We will show that when the magnetization is suppressed the SU(2) Schwinger representation is replaced by an U(1) spinon representation.

We will show this using the (radial) "particle-phase" representation:

$$b_\sigma^+(\vec{r}) = \sqrt{n_\sigma(\vec{r})}\, e^{-i\theta_\sigma(\vec{r})} = \sqrt{\frac{n(\vec{r}) + \hat{\sigma}n_s(\vec{r})}{2}}\, e^{-i\theta(\vec{r})}e^{-i\hat{\sigma}\tilde{\theta}_s(\vec{r})}$$
$$b_\sigma(\vec{r}) = e^{i\theta_\sigma(\vec{r})}\sqrt{n_\sigma(\vec{r})} = e^{i\theta(\vec{r})}e^{i\hat{\sigma}\tilde{\theta}_s(\vec{r})}\sqrt{\frac{n(\vec{r}) + \hat{\sigma}n_s(\vec{r})}{2}}\ .$$
(4)

In eq (2) we use the convention; $\hat{\sigma}(\sigma = \uparrow) = 1$ and $\hat{\sigma}(\sigma = \downarrow) = -1$.

In eq. (2) we have the commutation relations; $[n_\sigma(\vec{r}), \theta_{\sigma'}(\vec{r})] = i\delta_{\sigma,\sigma'}$. The bosonic pairs $n_\sigma(\vec{r})$, $e^{i\sigma\theta_\sigma(\vec{r})}$ are replaced in terms of the "density" and "spinon" bosons; $n_\sigma(\vec{r}) = \frac{1}{2}(n(\vec{r}) + \hat{\sigma}n_s(\vec{r}))$, $\theta_\sigma(\vec{r}) = \theta(\vec{r}) + \hat{\sigma}\tilde{\theta}_s(\vec{r})$ with the commutation relations; $[n(\vec{r}), \theta(\vec{r})] = i$, $[n_s(\vec{r}), \theta_s(\vec{r})] = i$. We replace, in the $b_\sigma(\vec{r})$, $b_\sigma^+(\vec{r})$ the density $r(\vec{r})$ by the average density $\bar{n} \equiv \langle n(\vec{r})\rangle = 1 - \langle\psi^+\psi\rangle \approx 1 - \delta$. As a result the bosonic fields are replaced by $e^{i\tilde{\theta}_s(\vec{r})}$, $e^{-i\tilde{\theta}_s(\vec{r})}$,

$$b_\uparrow^+(\vec{r})e^{i\theta(\vec{r})} \approx \frac{1}{\sqrt{2}}\left[\sqrt{\bar{n}(\vec{r})}\, e^{-i\tilde{\theta}_s(\vec{r})} + \frac{n_s(\vec{r})}{2\sqrt{\bar{n}(\vec{r})}}e^{-i\tilde{\theta}_s(\vec{r})}\right]$$
$$b_\downarrow^+(\vec{r})e^{i\theta(\vec{r})} \approx \frac{1}{\sqrt{2}}\left[\sqrt{\bar{n}(\vec{r})}\, e^{i\tilde{\theta}_s(\vec{r})} - \frac{n_s(\vec{r})}{2\sqrt{\bar{n}(\vec{r})}}e^{i\tilde{\theta}_s(\vec{r})}\right]$$
(4a)

and

$$e^{-i\theta(\vec{r})}b_\uparrow(\vec{r}) \approx \frac{1}{\sqrt{2}}\left[e^{i\tilde{\theta}_s(\vec{r})}\sqrt{\bar{n}(\vec{r})} + e^{i\tilde{\theta}_s(\vec{r})}\frac{n_s(\vec{r})}{2\sqrt{\bar{n}(\vec{r})}}\right]$$

$$e^{-i\theta(\vec{r})}b_\downarrow(\vec{r}) \approx \frac{1}{\sqrt{2}}\left[e^{-i\tilde{\theta}_s(\vec{r})}\sqrt{\bar{n}(\vec{r})} - e^{-i\tilde{\theta}_s(\vec{r})}\frac{n_s(\vec{r})}{2\sqrt{\bar{n}}}\right].$$

(4b)

The set of equations 4(a,b) have the property that in the limit $n_s(\vec{r})/n(\vec{r}) \ll 1$, the SU(2) bosonic fields $b_\sigma(\vec{r})$, $b_\sigma^+(\vec{r})$ can be replaced by the U(1) spinon, $e^{\pm i\tilde{\theta}_s(\vec{r})}$. The spinon field is related to the SU(2) fields, $b_\downarrow^+(\vec{r})e^{i\theta(\vec{r})} = b_\downarrow(r)e^{-i\theta(\vec{r})} \approx \sqrt{\bar{n}(r)}e^{-i\tilde{\theta}_s(\vec{r})}$;

$b_\downarrow^+(\vec{r})e^{+i\theta(\vec{r})} = b_\downarrow(\vec{r})e^{-i\theta(\vec{r})} \approx \sqrt{\bar{n}(r)}e^{-i\tilde{\theta}_s(\vec{r})}$.

Next we absorb the phase $\theta(r)$ into the fermion fields;

$$\psi(\vec{r}) = e^{-i\theta(\vec{r})}\tilde{\psi}(\vec{r})$$
$$\psi^+(\vec{r}) = \tilde{\psi}^+(\vec{r})e^{i\theta(\vec{r})}.$$

(5a)

The constraint in terms of the new fields is:

$$n_s(\vec{r}) = \sigma(\vec{r})\left(1 - \psi^+(\vec{r})\psi(\vec{r})\right).$$

(5b)

where $\sigma(\vec{r}) = \pm 1$.

Eqs (4a-5b) allow us to express the Hamiltonian given in eq. (1) in terms of the spinons and holons fields: $\theta_s(\vec{r})$, $n_s(\vec{r})$ and $\psi(\vec{r})$, $\psi^+(\vec{r})$.

4. THE HAMILTONIAN IN THE NEW REPRESENTATION-FERMIONS HOLONS INTERACTING WITH THE BOSONIC SPINON.

The Hamiltonian takes the form

$$H = \sum_{(\vec{r})} h(\vec{r}),$$
$$h(\vec{r}) = h_t(\vec{r}) + h_{t'}(\vec{r}) + h_\mu(\vec{r}) + h_c(\vec{r}) + h_m(\vec{r}) + \delta h_t(\vec{r}) + \delta h_{t'}(\vec{r}).$$
(6)

Here $h_t(\vec{r})$ is the nearest neighbor hopping term,

$$h_t(\vec{r}) = -\tilde{t} \sum_{(\vec{r})} \psi^+(\vec{r} + a_i)[\cos(\theta_s(\vec{r}) - \theta_s(\vec{r} + a_i)]\psi(\vec{r}) + \text{H.c.}$$
(6a)

"$h_{t'}$" is the next-nearest neighbor hopping.

$$h_{t'}(\vec{r}) = -\tilde{t}'/2 \sum_{t,-} \psi^+(\vec{r} + a_i + pa_2)[\cos(\theta_s(\vec{r}) - \theta_s(\vec{r} + a_i + pa_2)]\psi(\vec{r}) + \text{H.c.}.$$
(6b)

where $\tilde{t} \equiv t\bar{n} \simeq t(1-\delta)$, $.\tilde{t}' \simeq t'(1-\delta)$ where $\delta \equiv \langle \psi^+(\vec{r})\psi(\vec{r}) \rangle$

The Heisenberg Hamiltonian $h_m$ takes the form:

$$h_m(\vec{r}) = J(1-2\delta)\left(\frac{\bar{n}}{2}\right)^2 \sum_{i=1,2} \left(1 - \left(\frac{n_s(\vec{r})}{\bar{n}}\right)^2\right)^{1/2} \left(1 - \left(\frac{n_s(\vec{r}+a_i)}{\bar{n}}\right)^2\right)^{1/2} \cos[2(\theta_s(\vec{r}) - \theta_s(\vec{r}+a_i))]$$
$$+ \frac{J}{4}(1-2\delta) \sum_{i=1,2} n_s(\vec{r}) r_s(\vec{r}+a_i).$$
(6c)

The first part of eq. (6c) represent the $X-Y$ interaction and the second term $n_s(\vec{r})n_s(\vec{r}+a_i)$ is the Ising term. "$h_c$" is the "effective" uniaxial term,

$$h_c(\vec{r}) = \frac{D}{2} n_s^2(\vec{r}) \; , \quad D > 0 \; . \tag{6d}$$

The uniaxial parameter is normalized by the fluctuations induced by the doping. As a result "D" is replaced by $D_R > D > 0$. In particular we mention that, due to doping, one expects that even in the limit $D \to 0$, $D_R$ will be finite causing the destruction of the magnetic order.

"$h_\mu$" is the charge part. It contains the attractive charge-charge interaction (the term proportional to $-\frac{J}{4} n(\vec{r})n(\vec{r}+a_i)$ in eq. (1) and a shift of the chemical potential):

$$h_\mu(\vec{r}) = \left(-\mu + \frac{3}{4}J\right)\psi^+(\vec{r})\psi(\vec{r}) - \frac{J}{4} \sum_{i=1,2} \psi^+(\vec{r})\psi^+(\vec{r}+a_i)\psi(\vec{r}+a_i)\psi(\vec{r}) \tag{6e}$$

"$\mu$" is the chemical potential chosen such that, $\langle \psi^+(\vec{r})\psi(\vec{r})\rangle = \delta$. "$\delta h_t$" and "$\delta h_{t'}$" are irrelevant terms in the Renormalization Group sense and will be ignored.

# 5. THE EUCLIDEAN ACTION OF THE MODEL - FERMIONIC HOLONS AND BOSONIC SPINONS

In the second step we perform a Legendre transform and replace the Hamiltonian in eq. (6) by the action S. The transform is implemented by using the fact that $\psi^+(\vec{r})$ is the canonical momentum conjugate to $\psi(\vec{r},\tau)$ and $n_s(\vec{r},\tau)$ is the canonical momentum conjugate to $\theta_s(\vec{r},\tau)$ ("$\tau$" is the Euclidean time, $\tau = it$). The action has to be supplemented with the constraints n, $n_s(\vec{r}) = \sigma(\vec{r})\left(1 - \psi^+(\vec{r})\psi(\vec{r})\right)$ where $n_s(r)$ is an integer restricted by the Ising field, $\sigma(\vec{r}) = \pm 1$.

We will show that our action is equivalent to fermionic holons coupled to Bosonic spinons. We will show this by deriving the Berry phase of the two excitations.

We do this by construct the partition function:

$$Z = T_r\left[P_\sigma P_{n_s} T_\tau \exp\left(-\int_0^\beta d\tau H(\tau)\right)\right]$$

$$= T_r T_\tau \prod_{\tau_n = a_r \beta/N}^\beta \left[P_\sigma(\tau_n) P_{n_s}(\tau_n) \exp[-\varepsilon H(\tau_n)]\right] , \qquad (7a)$$

where $P_\sigma(\tau_n)$ and $P_{n_s}(\tau_n)$ are given by:

$$P_\sigma(\tau_n) = \prod_{\vec{r}} \int \frac{d\lambda(\vec{r},\tau_n)}{2\pi} \exp\left\{i\lambda(\vec{r},\tau_n)\left[n_s(\vec{r},\tau_n) - \sigma(\vec{r},\tau_n)\left(1 - \psi^+(\vec{r},\tau_n)\psi(\vec{r},\tau_n)\right)\right]\right\}$$

(7b)

and

$$P_{n_s}(\tau_n) = \prod_{\vec{r},\tau_n} e^{i2\pi n_s(\vec{r},\tau_n)m_o(\vec{r},\tau_n)} ,$$

(7c)

where $-\infty \leq m_o \leq \infty$ are integers.

Eq. (7a) is evaluated using the "Bosonic" coherent states $|\theta_s(\vec{r},\tau)\rangle$ and the "Grassman" coherent states $|-\overline{\psi}(r,\tau)\rangle, |\psi(\vec{r},\tau)\rangle$. Using these coherent states we compute the trace in eq. (7a):

$$T_r[\ ] \equiv \sum_{\sigma=\pm 1} \sum_{m_o=-\infty}^{\infty} \iint \langle -\overline{\psi},\theta_s|[\ ]|\theta_s,\psi\rangle e^{-\overline{\psi}\psi} d\overline{\psi}d\psi d\theta_s dn_s .$$

(7d)

We sum over $\sigma = \pm 1$ and find the partition function Z:

$$Z = \int e^{-S} d\overline{\psi}d\psi d\theta_s dn_s d\lambda .$$

(7e)

The summation over "$m_o$" is taken care of by replacing the range of integration of $\theta_s$ from $[+\pi,\pi]$ to $[-\infty,\infty]$.

The action in eqs. (7e,8) has two parts : the Berry phase, the Hamiltonian and the coupling between holons and spinons generated by the field $\lambda$ .

$$S = \sum_{\vec{r}} \sum_{\tau} \Big\{\overline{\psi}(\vec{r},\tau)\hat{\partial}_\tau \psi(r,\tau) - in_s(\vec{r},\tau)\hat{\partial}_\tau \theta(\vec{r},\tau)$$
$$- i\lambda(\vec{r},\tau)n_s(\vec{r},\tau) - \log|\cos(\lambda(\vec{r},\tau)(1-\psi^+(\vec{r},\tau)\psi(\vec{r},\tau))|+a_o h(\vec{r},\tau)\Big\} .$$

(8)

In eq. (8), the first term is the fermionic Berry phase for the holons and the second term is the Bosonic Berry phase for the spinons.

Additional parameters in the action are, $a_o" = \frac{\beta}{N}$ is the "temporal" lattice constan, "$\hat{\partial}$" is the lattice derivative, $\hat{\partial}_\tau f \equiv f(\tau + a_i) - f(\tau)$, ($\hat{\partial}_\tau f$ is related to the regular time derivative $\partial_\tau \equiv \hat{\partial}_\tau / a_o$).

In eq. (7e) we perform the trace over $\sigma = \pm 1$. As a result the term "$i\lambda n_s$" in eq. (8) is replaced by the term $-\log|\cos\lambda(\vec{r},\tau)| = \frac{\lambda^2}{2} + \frac{\lambda^4}{12} \cdots$ .

Next we expand the magnetic Hamiltonian around the antiferromagnetic/ground state. We introduce the new fields $\hat{\theta}_s(\vec{r},\tau)$, and $\hat{n}_s(\vec{r},\tau)$,

$$n_s(\vec{r},\tau) = (-1)^{(r_x + r_y)} \hat{n}_s(\vec{r},\tau) \tag{9a}$$

$$\theta_s(\vec{r},\tau) = \frac{\pi}{4}(-1)^{(r_s + r_y)} + \hat{\theta}_s(\vec{r},\tau)$$

(9b)

where $r_x + r_y$ is an even/odd integer. Next we integrate the field $\lambda(\vec{r},\tau)$ and find to order $n_s^4(\vec{r})$ the effective action,

$$S = \sum_{\vec{r}} \sum_{\tau} \left\{ \overline{\psi}(\vec{r},\tau)\hat{\partial}_{\tau}\psi(\vec{r},\tau) - i(-1)^{r_x+r_y}\hat{n}_s(\vec{r},\tau)\hat{\partial}_{\tau}\hat{\theta}_s(\vec{r},\tau) + \frac{1}{2}\left(1+\psi^+(\vec{r},\tau)\psi(\vec{r},\tau)\right)\hat{n}_s^2(\vec{r},\tau) \right.$$

$$+ \frac{1}{12}\left(1+\psi^+(\vec{r},\tau)\psi(\vec{r},\tau)\right)\hat{n}_s^4(\vec{r},\tau)$$

$$+ a_o \left[ \frac{D}{2}\hat{n}_s^2(\vec{r},\tau) - \frac{J_\parallel}{4}\sum_{i=1,2}\hat{n}_s(\vec{r},\tau)\hat{n}_s(\vec{r}+a_i,\tau) - \frac{J_\perp}{4}\sum_{i=1,2}\cos(2\hat{\partial}_i\hat{\theta}_s(\vec{r},\tau)) \right.$$

$$-\tilde{t}\sum_{i=1,2}\overline{\psi}(\vec{r}+a_1,\tau)\sin(\hat{\partial}_i\hat{\theta}_s(\vec{r},\tau))\psi(\vec{r},\tau) + \text{H.c.}$$

$$-\tilde{t}'\sum_{p=+,-}\overline{\psi}(\vec{r}+a_1+pa_2)\cos(\hat{\partial}_1\hat{\theta}_s(\vec{r},\tau)+p\hat{\partial}_2\hat{\theta}_s(\vec{r},\tau))\psi(\vec{r},\tau) + \text{H.c.}$$

$$\left. \left. + h_\mu(\vec{r},\tau) + \delta h_t(\vec{r},\tau) + \delta h_{t'}(\vec{r},\tau)) \right] \right\}.$$

(10)

In eq. (10) we used the lattice derivative $\hat{\partial}_i = a\partial_i$ ("$a$" is the lattice constant). The exchange parameters $J_\perp$ and $J_\parallel$ are given by:

$$J_\perp = J(1-2\delta)\overline{n}^2\left(1 - \frac{\langle \hat{n}_s^2 \rangle}{\overline{n}^2}\right)$$

$$J_\parallel = J(1-2\delta).$$

The Ising term in eq. (10), takes the form

$$\frac{J_\parallel}{4}\sum_{\vec{r}}\sum_{i=1,2}\hat{n}_s(\vec{r},\tau)\hat{n}_s(\vec{r}+a_i,\tau) = \frac{J_\parallel}{4}\sum_{\vec{k}}\hat{n}_s(\vec{k},\tau)(2\cos(k_x a)+2\cos k_y a)\hat{n}_s(-\vec{k},\tau)$$

$$\approx J_\parallel\sum_{\vec{k}}\hat{n}_s(\vec{k},\tau)\hat{n}_s(-\vec{k},\tau) - \frac{J_\parallel}{4}\sum_{\vec{k}}|\vec{k}|^2\hat{n}_s(k,\tau)\hat{n}_s(-\vec{k},\tau).$$

(11)

Combining the $\vec{k} = 0$ term from eq. (11) with the term induced by fluctuations, we obtain the renormalized uniaxial term, $\frac{D_R}{2} \hat{n}_s^2(\vec{r},\tau)$. D is replaced by $D_R$:

$$\left[\left(1+\langle \psi^+\psi\rangle\right)\left(1+\frac{1}{12}\langle n_s^2\rangle\right) - 2J_\| a_0 + Da_0\right] \equiv \frac{1}{2} a_0 D_R \qquad (12)$$

$D_R = 0$ corresponds to a critical concentration $\langle \overline{\psi}\psi\rangle = \delta_c$,

$$\delta_c = \langle \overline{\psi}\psi\rangle = \left[(2J_\| - D)a_0 - \left(1+\frac{1}{12}\langle n_s^2\rangle\right)\right]\frac{1}{1+\frac{1}{12}\langle n_s^2\rangle} .$$

For $\delta > \delta_c$, $D_R > 0$ and $<n_s> = 0$. We observe that the uniaxial term $D_R$ is controlled by the doping $\langle \overline{\psi}\psi\rangle$ and the quantum fluctuations of the $\hat{n}_s^4(\vec{r},\tau)$ term.

6. THE NON MAGNETIC PHASE –FERMIONIC HOLONS INTERACTING WITH THE SPINONS OR WITH THE MONOPOLES CURRENTS

For the remaining discussion we consider <u>only</u> the situation where the uniaxial term $D_R$ is positive such that $\hat{n}_s = 0$ and the fluctuations of ($\hat{n}_s$) are small such that $J_\perp \neq 0$. When $D_R > 0$ we can integrate the "canonical momentum" $\hat{n}_s(\vec{r},\tau)$ and generate an action in terms of $\partial_\mu \hat{\theta}_s(\vec{r},\tau)$ and $\hat{\theta}_s(\vec{r},\tau)$ only. In the continuous limit we obtain:

$$S = \int d^2r \int d\tau \left\{ \overline{\psi}(\vec{r},\tau)\partial_\tau \psi(\vec{r},\tau) + \frac{1}{2}\gamma_o(\partial_\tau \hat{\theta}_s(\vec{r},\tau))^2 - \frac{1}{2}\gamma_\perp \sum_{i=1,2} a^{-2} \cos(2\hat{\partial}_i \hat{\theta}_s(\vec{r},\tau)) \right.$$

$$-\tilde{t} \sum_{i=1,2} \overline{\psi}(\vec{r}+a_i,\tau)[\sin\hat{\partial}_i \hat{\theta}_s(\vec{r},\tau)]\psi(\vec{r},\tau) + \text{H.c.}$$

$$-\tilde{t}' \sum_{p=+,-} \overline{\psi}(\vec{r}+a_1+pa_2,\tau)\left[\cos(\hat{\partial}_1 \hat{\theta}_s(\vec{r},\tau)+p\hat{\partial}_2 \hat{\theta}_s(\vec{r},\tau))\right]\psi(\vec{r},\tau) + \text{H.c.} + h_\mu(\vec{r},\tau) \right\} \quad (13)$$

$$\equiv S^{(x-y)} + S^{(\psi)} .$$

In eq. (11) we have neglected the term $\delta h_t$ and $\delta h_{t'}$. This is justified for $D_R > 0$.

In eq. (11) we have defined $\gamma_o \equiv \frac{1}{D_R}$ and $\gamma_\perp \equiv \frac{J_\perp}{2}$, and we observe that the spinons fluctuations couple to the holoons (charge fluctuations) through the hopping term. The major difference in eq. 13 with the U(1) gauge theory is the fact that the gauge field appears as a "real field" and not as a complex order parameter. The "cosine" term controls the gauge field fluctuations. As a result of this eq. 13 is only invariant for "discreet" $Z_2$ transformation.

We observe that when we perform a gauge transformation, $\psi \to \psi e^{i\alpha}$, $\overline{\psi} \to \overline{e}^{i\alpha}\psi$, $\hat{\theta} \to \hat{\theta} + \alpha$. Equation 11 remains invariant only if the phase $\alpha(\vec{r},\tau)$ is given by $\alpha(\vec{r},\tau) = \pi m(\vec{r},\tau)$ with "m" being integers. For $m = 1,-1$ we have a $Z_2$ gauge model with the phase fluctuation $\pi m(\vec{r},\tau)$ representing the vortex excitation of $\cos[2\hat{\partial}_i \hat{\theta}_s]$. The hamiltonian in eq. 13 can be analyzed according to the the spinon phase fluctuations $\hat{\theta}_s$:

a) We have a spinon (spin-wave phase) with smal fluctuations . In this phase the "cosine" terms are replaced by their expectation values. As a result we obtain a spinon-holons (spin-charge) separated phase.

b) A vortex monopole current phase, in this phase fluctuations are large and no spin-charge separation is possible.

The phase transition between the two phases gives rise in eq. 13 to $T = 0$ to a superconducting non-superconducting transition. The vortex loop phase is not superconducting at $T = 0$ and describe at $T \neq 0$ a regular metals. The vortex free phase at $T = 0$ describes a superconductor and becomes at finite temperature a non-fermi-liquid.

From eq. 13 we conclude that the <u>spin-charge coupling is induced by the vortex excitation.</u> Therefore a spin-charge separated phase is possible in the absence of the vortex excitations.

## 7. THE SPIN CHARGE SEPARATION CAUSED BY THE FREEZING OF THE MONOPOLE CURRENTS

In order to investigate the possible phases of eq. (13) we consider first the $X - Y$ action controlled by the spinon order parameter $\hat{\theta}_s(\vec{r}, \tau)$:

$$S^{(x-y)} = \int dr^2 \int d\tau \left[ \frac{1}{2} \gamma_0 \cdot \left( \partial_\tau \hat{\theta}_s(\vec{r}, \tau) \right)^2 - \frac{1}{2} \gamma_\perp \sum_{i=1,2} a^{-2} \cos\left( 2\hat{\partial}_i \hat{\theta}_s(\vec{r}, \tau) \right) \right] . \qquad (14a)$$

Eq. (14a) represents the $X-Y$ model in $2+1$ dimensions which is equivalent to a gas of "monopole current loops", $\ell_\mu(\vec{r},\tau)$, $\mu=0,1,2$ which satisfy, $\partial_\mu \ell_\mu = 0$. Following Bank and Meyerson (12) we replace the $S^{(x-y)}$ action with the current loop action, $S^{(\text{loop})}$:

$$S^{(\text{loop})} = \frac{\pi^2 \gamma_{\text{eff}}}{2} \int dr^2 \int d^2 r' \int d\tau \int d\tau' \sum_{\mu=0,1,2} \ell_\mu(\vec{r},\tau) V(\vec{r}-\vec{r}',\tau-\tau') \ell_\mu(\vec{r}',\tau') , \qquad (14b)$$

where $\gamma_{\text{eff}} \equiv \gamma_\perp \left(\dfrac{\gamma_o}{\gamma_\perp}\right)^{1/2}$ is the effective coupling and the loop currents obey the conservation law $\partial_\mu \ell_\mu = 0$. The potential $V(\vec{r}-\vec{r}',\tau-\tau')$ is given in the strong coupling limit by the unscreened potential, $V = \dfrac{1}{\sqrt{(\vec{r}-r')^2 + (\tau-\tau')^2}}$.

Following Ref. 12, we find that the model in eq. (14b) has a phase transition at $\gamma_{\text{eff}} \simeq \gamma^* \simeq \dfrac{8}{6.3} \simeq 1.25$. For $\gamma_{\text{eff}} < \gamma^*$ the current loops are in the plasma phase giving rise to a screened potential $V(\vec{r},\vec{r}',\tau-\tau') \sim V(0)\delta_{\vec{r},\vec{r}'}\delta_{\tau,\tau}$. This is the phase of free monopoles currents. On the other hand, for $\gamma_{\text{eff}} < \gamma^*$ the monopoles currents are not free and the potential is the unscreened Coulomb potential.

In order to compute the properties of the holons ( charge holes) we have to calculate expectation value of the term $\sin(\hat{\partial}_i \hat{\theta})$, and $\cos[\hat{\partial}_1 \hat{\theta}_S + p\hat{\partial}_2 \hat{\theta}_S]$, $p = +,-$ in eq. (13). The expectation values will be computed using the monopole currents loops.

We do this by expressing the large fluctuations of $\partial_i \hat{\theta}_S$ by a vector potential, $A_i = (\partial_i \hat{\theta}_S)_v$. This vector potential describes the vorticity. Expressing the vector potential $A_i, i = 1,2$. In terms of the vortex loops $\ell_i$ $i = 1,2$ we find:

$$\left(\partial_i \hat{\theta}_S(\vec{r},\tau)\right)_v = \frac{1}{4\pi}\varepsilon_{ij}\int d^2 r' \int \frac{d\tau'}{\sqrt{(\vec{r}-\vec{r}')^2 + (\tau-\tau')^2}} \ell_J(\vec{r}',\tau')\sin[\alpha(\vec{r}-\vec{r}',\tau-\tau')] \ . \quad (14c)$$

Here "$\alpha$" is the polar angle between the current loop and the unit vector, $\frac{(\vec{r}-\vec{r}',\tau-\tau')}{\sqrt{(\vec{r}-\vec{r}')^2 + (\tau-\tau')^2}}$. Using eq. (14c) and eq. (10b) we find that, $\left\langle e^{i(\partial_i \hat{\theta}_S)_v} \right\rangle$ is nonzero for the unscreened potential ($\gamma_{eff} > \gamma^*$), and zero for the screened potential $\gamma_{eff} < \gamma^*$. As a result, for $\gamma_{eff} > \gamma^*$ we have spin-charge separation and for $\gamma_{eff} < \gamma^*$ the vortex free phase will give rise to the non-separated spin-charge phase.

Next we concentrate our discussion on the spin-charge separated phase, $\gamma_{eff} > \gamma^*$. In this region we replace $t'$ by $t_{eff}$. As a result eq. (13) is replaced by

$$S = S_{eff}^{(\psi)} + S_{spin-wave}^{(x-y)} \quad (15a)$$

$$S_{eff}^{(\psi)} = \int d^2 r \int d^2 \tau \{\overline{\psi}(r,\tau)(\partial_\tau - \hat{\mu})\psi(r,\tau)$$
$$-t_{eff}\sum_{p=+,-}\overline{\psi}(\vec{r}+a_1+pa_2,\tau)\psi(\vec{r},\tau) + H.c. - \frac{J}{4}\sum_{i=1,2}\overline{\psi}(\vec{r})\psi(\vec{r})\overline{\psi}(\vec{r}+a_i)\psi(\vec{r}+a_i) \quad (15b)$$

and the spin-wave action is

$$S^{x-y}_{spin-wave} = \int d^2r \int d\tau \left[ \frac{1}{2}\gamma_o(\partial_\tau \hat{\theta}_s)^2 - \frac{1}{2}\tilde{\gamma}_\perp \sum_{i=1,2}(\partial_i \hat{\theta}_S)^2 \right]; \qquad \tilde{\gamma}_\perp = 4\gamma_\perp , \qquad (15c)$$

where $t_{eff}$ is given by;

$$t_{eff} = t'\left\langle \cos(\partial_1\hat{\theta}_S + P\partial_2\theta_S) \right\rangle_{spin-wave} > \neq 0 . \qquad (15d)$$

The presence of the attractive interaction in eq. (15b) gives rise at low temperatures to superconductivity in the spin-charged separated phase. The effective model in eq 15b is controlled by the nearest neighbor pairing energy $J/4$ and the effective hopping term $t_{eff}$ (the nearest neighbor hopping vanishes in the spin-wave phase). Therefore the normal state will exist only for $T > T_c(J)$. $T_c(J)$ is the superconducting critical temperature for the model in eq. 15b. The physics at zero temperatures is determined by the effective coupling constant. From the analysis of eq. (14b), we know that $\gamma_{eff} = \gamma^*$ is the quantum critical point. Since $\gamma_{eff}$ is a function of "$\delta$" (see eq. (12)) we conclude that the quantum critical point $\gamma^*$ is controlled by doping, $\gamma^* \equiv \gamma_{eff}(\delta = \delta^* > \delta_c)$ describes the zero temperature the superconducting phase transition . For $\gamma_{eff} < \gamma^*$ the superconducting ground state is destroyed by the free monopoles currents.

8.CONCLUSION

We conclude by summarizing our main result. Using a bosonic formulation we have constructed an effective action for the t-t'-J model in the presence of a positive uniaxial anisotropy. For holes concentrations $\delta$, $\delta > \delta_c$ the Ising magnetic order is destroyed. As a result the $t - t' - J$ model with the uniaxial anysotropy is replaced by a $t - t' - J_\perp$ model which has $X - Y$ symmetry. As a result the holes experience a gauge field with a $Z_2$ symmetry( and not a U(1) symmetry ).The spin fluctuations are thus described within the $X - Y$ model which has two phases — a spin-wave phase for, $\gamma_{eff} > \gamma^* \equiv \gamma(\delta = \delta^*)$ and vortex monopole loop phase for $\gamma_{eff} < \gamma^*$. Therefore, for $\delta_c < \delta \leq \delta^*$ we obtain a spin-charge (spino-holon) separated phase which is superconducting for temperatures T , $T < T_c(J)$ . The point $\gamma^* = \gamma(\delta^*)$ represent the quantum critical point. For $\delta > \delta^*$ we have a vortex liquid in this phase; spin-charge separation and superconductivity are both destroyed. As a result we find a superconducting –non-superconducting phase transition at T=0.